%% file: acnm_revised.tex
\documentstyle[epsfig,12pt]{elsart}

\begin{document}
\newcommand{\p}{\partial}
\newcommand{\ls}{\left(}
\newcommand{\rs}{\right)}
\newcommand{\beq}{\begin{equation}}
\newcommand{\eeq}{\end{equation}}
\newcommand{\beqa}{\begin{eqnarray}}
\newcommand{\eeqa}{\end{eqnarray}}
\begin{frontmatter}
\title{Asymmetric colliding nuclear matter approach in heavy ion collisions}
\author[catania]{T. Gaitanos},
\author[tuebingen]{C. Fuchs},
\author[muenchen]{H.~H. Wolter}
\address[catania]{Laboratori Nazionali del Sud INFN, I-95123 Catania, Italy } 
\address[tuebingen]{Institut f\"ur Theoretische Physik der 
Universit\"at T\"ubingen, D-72076 T\"ubingen, Germany} 
\address[muenchen]{Dept. f\"ur Physik, Universit\"at M\"unchen, 
D-85748 Garching, Germany}  
\begin{abstract}
The early stage of a heavy ion collision is governed by local non-equilibrium 
momentum distributions which have been approximated by colliding nuclear 
matter configurations, i.e. by two Lorentz 
elongated Fermi ellipsoids. This approach has been extended from the 
previous assumption of symmetric systems to asymmetric 2-Fermi 
sphere configurations, i.e. to different densities. This provides a 
smoother transition from the limiting situation of two interpenetrating 
currents to an equilibrated system. 
The model is applied to the dynamical situations of 
heavy ion collisions at intermediate energies within the framework 
of relativistic transport (RBUU) calculations. We find that the 
extended colliding nuclear matter approach is more appropriate 
to describe collective reaction dynamics in terms of flow observables, 
in particular, for the elliptic flow at low energies.
\end{abstract}

\end{frontmatter}
\section{Introduction}
Heavy ion collisions from Fermi energies up to 
relativistic energies of $1-2$ AGeV provide the unique possibility to explore 
the nuclear matter equation of state (EOS) under extreme conditions of 
density and temperature in the laboratory \cite{fopir}. 
In a hydrodynamical picture the time evolution of such a
reaction can be understood in terms of a pressure gradient which builds
up in the compressed zone and drives the dynamics.  
 Naively,  one therefore expected 
to obtain  a direct access to the nuclear EOS in the 
measurement of 
collective flow observables. However, in fact, the situation 
is more complex: The system does not behave like an ideal
fluid but binary nucleon-nucleon collisions lead to a viscous
behavior. Moreover, calculations show that over most of the reaction time, 
in particular
during the compressional phase, the system is out of {\it local} 
equilibrium, see e.g. \cite{temp}. Also experimental evidence for 
non-equilibrium in terms of 
incomplete stopping even in central reactions has recently been reported
\cite{fopi3}. Thus the hydrodynamical limit is not reached in
relativistic heavy ion reactions except in the final stages of the collision.

On the other hand, microscopic transport models have proved to be 
an adequate tool for the description of the non-equilibrium 
reaction dynamics at
intermediate energies \cite{fopir}. The physical 
input of such semi-classical models based on Boltzmann type equations 
are the nuclear mean field $U$ and the nucleon-nucleon (NN) cross section
$\sigma$.  Both are derived from the effective two-body interaction in the medium,
i.e. the in-medium G-matrix;  $U\sim (\Re G\rho),~~
\sigma\sim (\Im G\rho)$, respectively $d\sigma/d\Omega\sim | G|^2$. 
However, in most practical applications phenomenological mean fields
are used. Adjusting the known bulk properties of nuclear matter 
around saturation one has to rely on extrapolations to supra-normal densities.
One can then try to constrain the models with the help of 
heavy ion reactions \cite{dani00,sahu98}. 

As mentioned above, in heavy ion collisions 
the system is far away from local equilibrium even during the high density phase 
which mainly governs the dynamics of the process. 
In a fully consistent treatment non-equilibrium effects should be 
considered self consistently within the framework of 
microscopic many-body theory and a 
dynamical description of the reaction, e.g.  by Boltzmann type 
transport equation \cite{horror}. 
An exact solution of the problem would require a consistent treatment of the 
transport equation and the microscopic structure equations,
e.g. within the framework of Dirac-Brueckner (DB) theory, 
for general non-equilibrium situations. Thus one would have to
determine the relativistic in-medium G-matrix not only for an 
equilibrated Fermi sphere, but for arbritrary momentum distributions, 
in principle also at finite temperatures. 
This is presently not possible without introducing further approximations.
The simplest approximation is the Local Density 
Approximation (LDA), which assumes a local spherical, sharp Fermi sphere, 
and thus essentially neglects non-equilibrium effects in the in-medium 
interactions. As a step towards non-equilibrium the
Colliding Nuclear Matter (CNM) model \cite{sehn} was introduced 
(which was also called the Local phase space Configuration Approximation, LCA),
which parametrizes the local momentum distribution by two Fermi spheres 
 with a finite relative velocity. 
We have shown  previously that the non-equilibrium effects included
in the CNM model modify the underlying EOS, in particular, the dynamical EOS 
effectively {\it seen} during a heavy ion collision is softened 
compared to ground state matter at equivalent densities \cite{gait03}. 
This affects the collective dynamics, e.g. in terms 
of observable flow signals \cite{gait99,gait01}. From these studies it was 
concluded that this type of non-equilibrium features should be
included at the mean-field level in transport analysis of heavy ion 
reactions in order to allow more reliable conclusions on the nuclear
matter EOS.

The Colliding Nuclear Matter (CNM) model has been extensively studied 
in \cite{sehn} and has been applied to intermediate energy heavy ion 
collisions in refs. \cite{gait03,gait99,gait01}. There it was 
shown that the consideration of non-equilibrium effects within the 
CNM approach essentially  changes the description of collective dynamics 
relative to a treatment in the local density approximation
(LDA). 
Differences of the two
approximations appeared in collective flow observables and were found 
to be of the same magnitude as those which arise due to a different, 
i.e.  soft or a hard, EOS. 

In this work we extend our previous studies by including more 
details in the approximations involved in the description of non-equilibrium 
situations within the CNM approach. 
In previous applications a symmetric configuration was assumed, i.e. 
two nuclear matter streams of equal (averaged) density. This should not 
be very realistic, particularly in peripheral collisions.
For this purpose the Asymmetric 
Colliding Nuclear Matter (ACNM) model 
is introduced here, which accounts also for a possible asymmetry of the 
Fermi momenta of the two Fermi fluids. This situation  arises   
in those regions where the tail of one nucleus penetrates into the 
interior of the other (see e.g. Fig. 1). 
In previous studies based on the CNM approximation we 
mainly considered symmetric colliding systems. A locally asymmetric 
two-stream situation is also more frequently encountered in
collisions of different mass systems. Such reactions 
will experimentally be studied more often in the future. 
Thus the extended approximation is a further step 
into the direction of ref. \cite{horror}, i.e. the attempt to 
describe the complex 
dynamical situation in heavy ion reactions as precisely as possible. 
The extension to ACNM is found to be
particularly important at low and intermediate energies when 
peripheral heavy ion collisions are considered. 

\section{Approximations in colliding nuclear matter}

As discussed in the introduction the treatment of non-equilibrium effects 
in dynamical transport descriptions 
cannot be done without any approximations due to its high complexity. 
In the CNM approach \cite{sehn} ground state 
DB results are extrapolated 
for idealized $2$-Fermi spheres or, covariantly, for $2$-Fermi ellipsoids, 
at zero temperature. 
This approximation constructs the 2-Fermi sphere configuration in a 
covariantly consistent way with respect to the effective masses, but
neglects the blocking in the intermediate propagator due to the 
second current. It was shown in \cite{sehn} that this approximation  
is reliable due to 
a moderate dependence of the DB self energy around the  Fermi momentum. 

In applications to heavy ion reactions the relevant approximations 
are  
(a) the zero-temperature limit and 
(b) the restriction to symmetric CNM distributions, i.e. 
same Fermi momenta of the two counter-streaming currents. 
To demonstrate these effects we show in Fig. 1 typical local momentum 
distributions as they are 
obtained from transport descriptions of central 
Au+Au reaction at $0.6$ AGeV beam energy 
at different locations in coordinate space (in the center of mass, and in 
beam direction) at the time just before the compression phase. In the 
left column we show the local momentum distributions as they are obtained
from relativistic BUU calculations. Details of the RBUU model used
here are given in section 4. 
In the right column are asymmetric fits to the
momentum distribution, i.e. fits with two Fermi distributions (including 
Pauli corrections) of different density and temperature, for details see 
\cite{temp}.   It is seen 
from these fits that the momentum space 
consists in general of two rather different Fermi distributions outside 
of the central cell, which are furthermore characterized
by different local temperatures of the subsystems.

\begin{figure}[t]
\begin{center}
\unitlength1cm
\begin{picture}(14,18)
\put(-4.5,12){\makebox{\epsfig{file=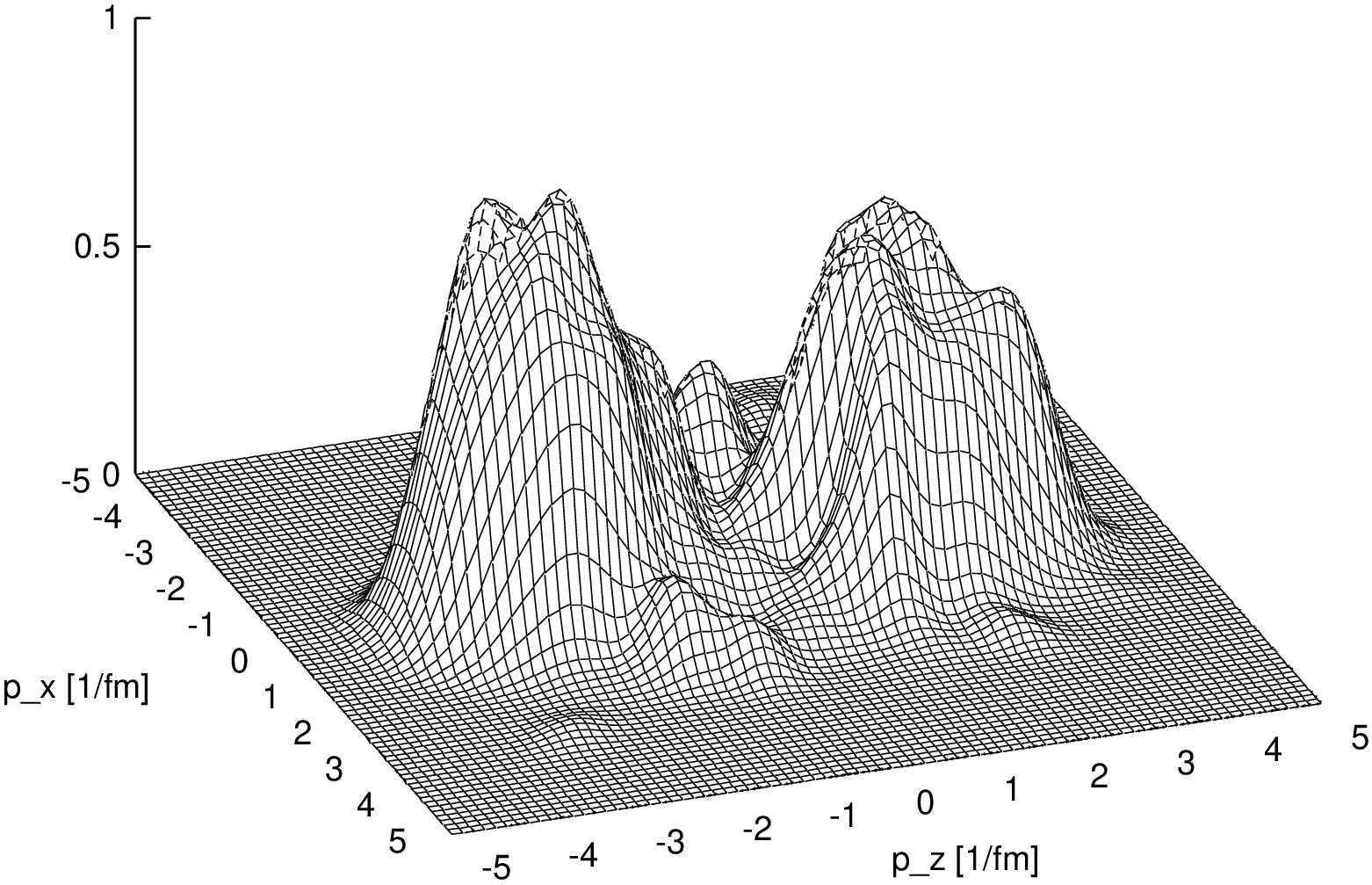,width=8.0cm}}}
\put(-4.5,6){\makebox{\epsfig{file=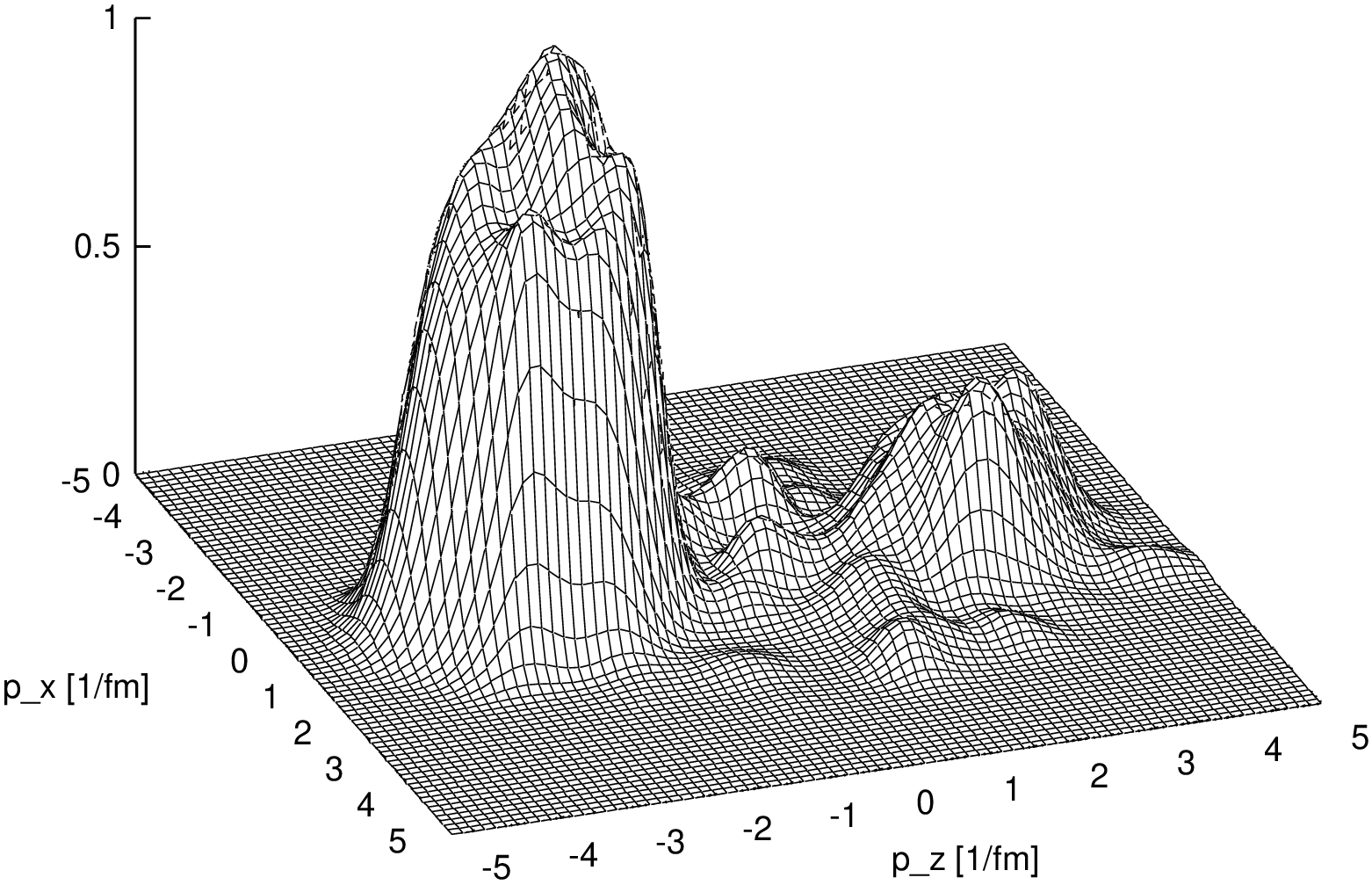,width=8.0cm}}}
\put(-4.5,0){\makebox{\epsfig{file=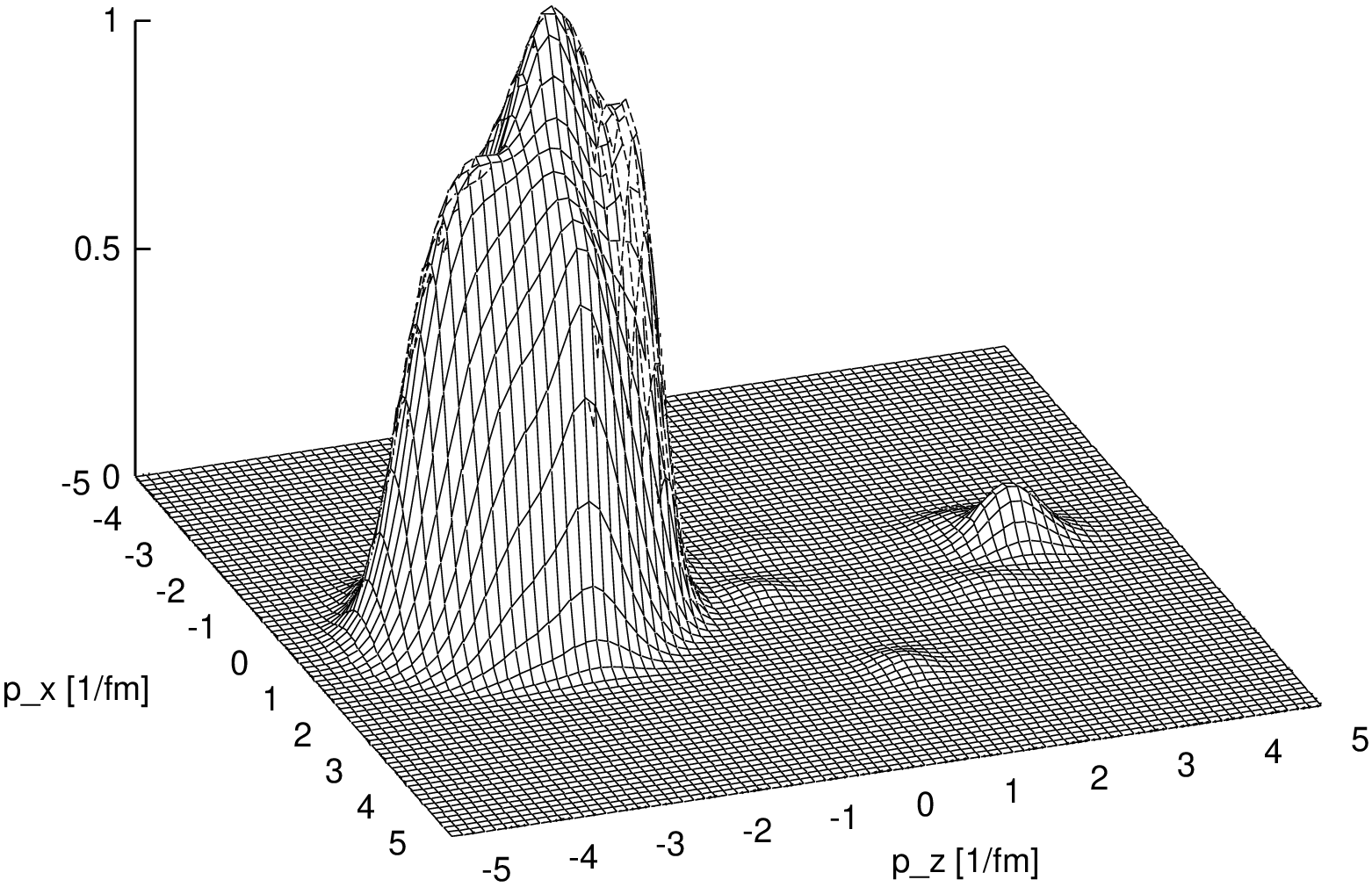,width=8.0cm}}}
\put(3.5,12){\makebox{\epsfig{file=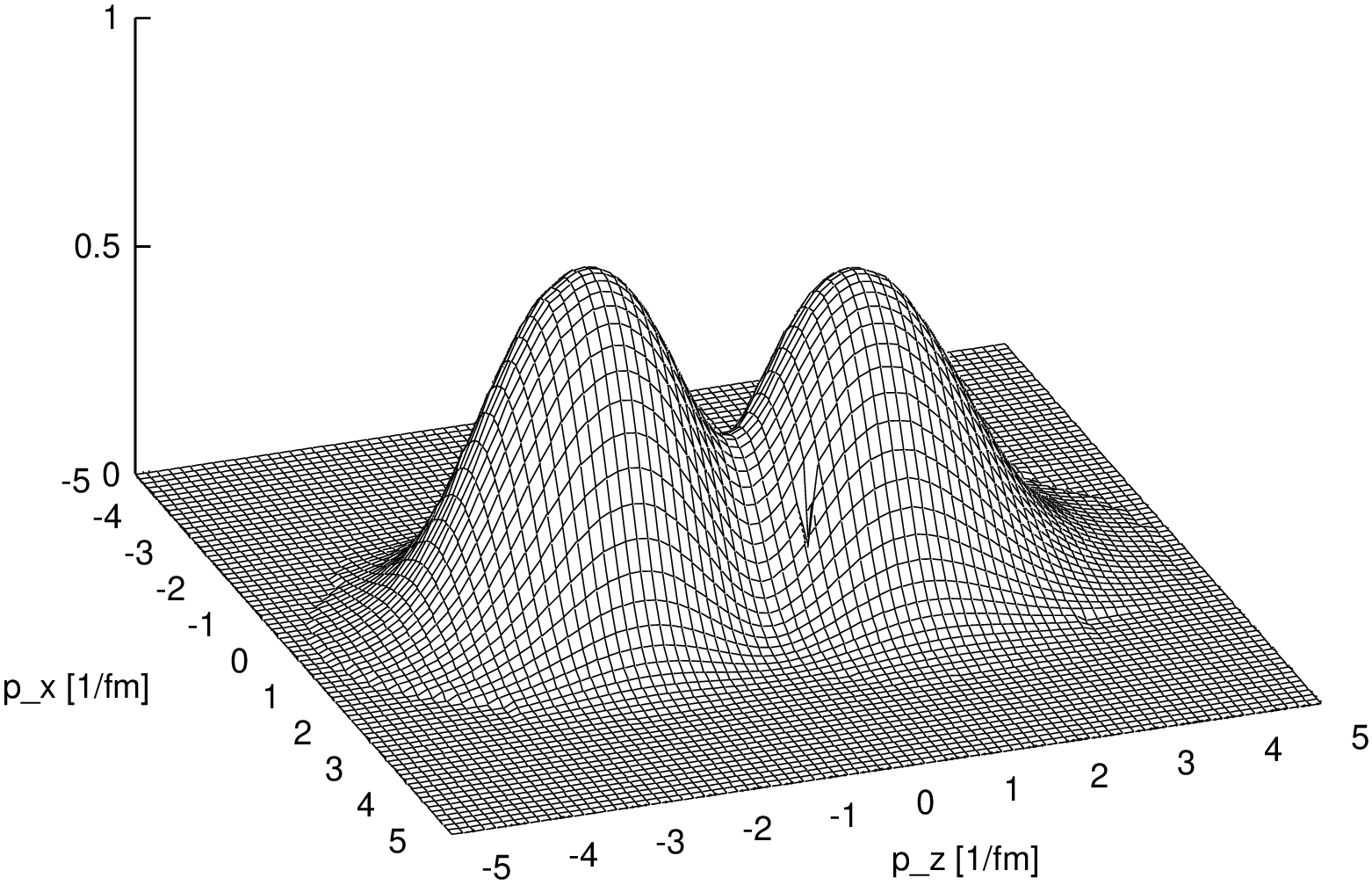,width=8.0cm}}}
\put(3.5,6){\makebox{\epsfig{file=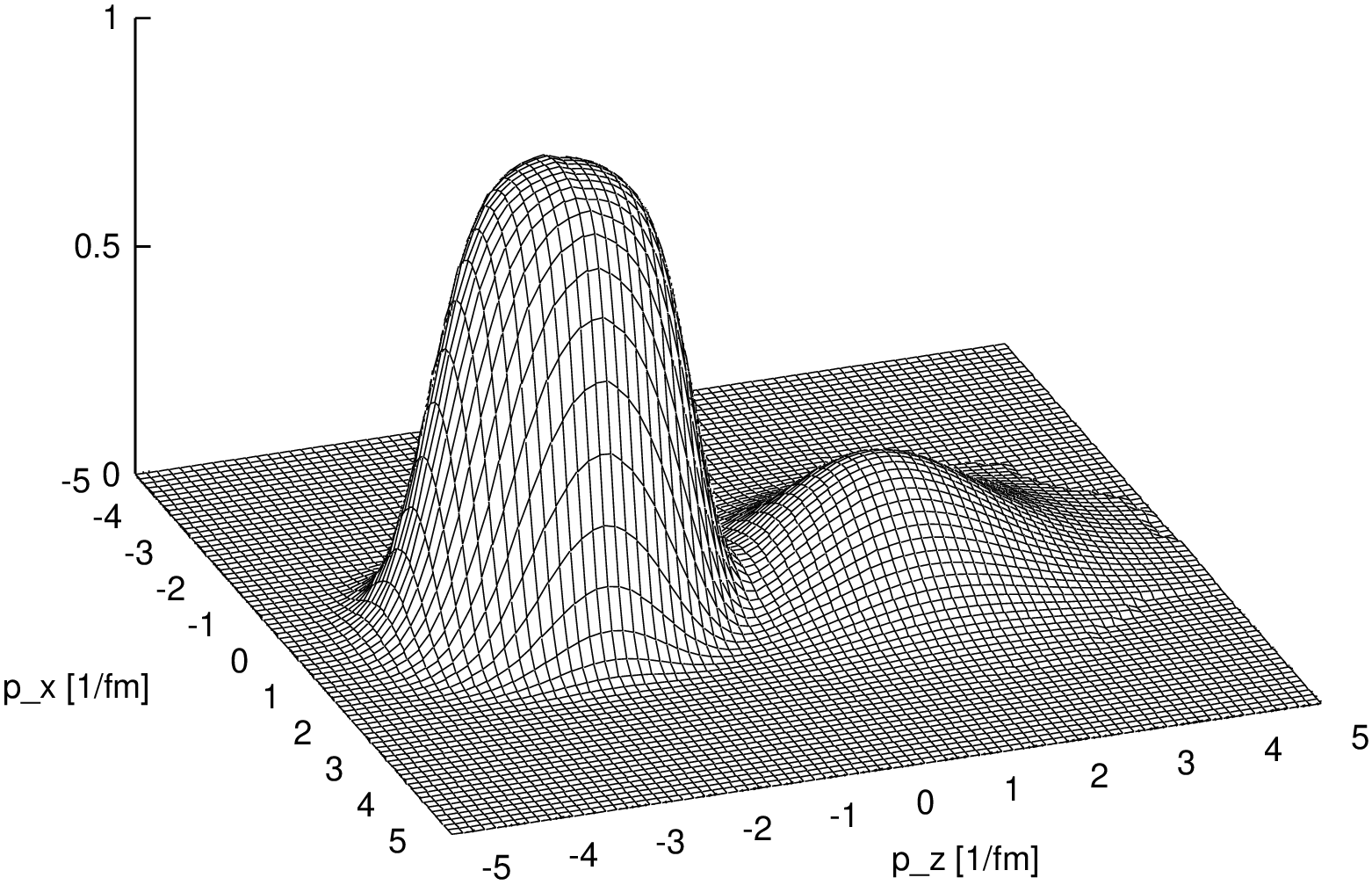,width=8.0cm}}}
\put(3.5,0){\makebox{\epsfig{file=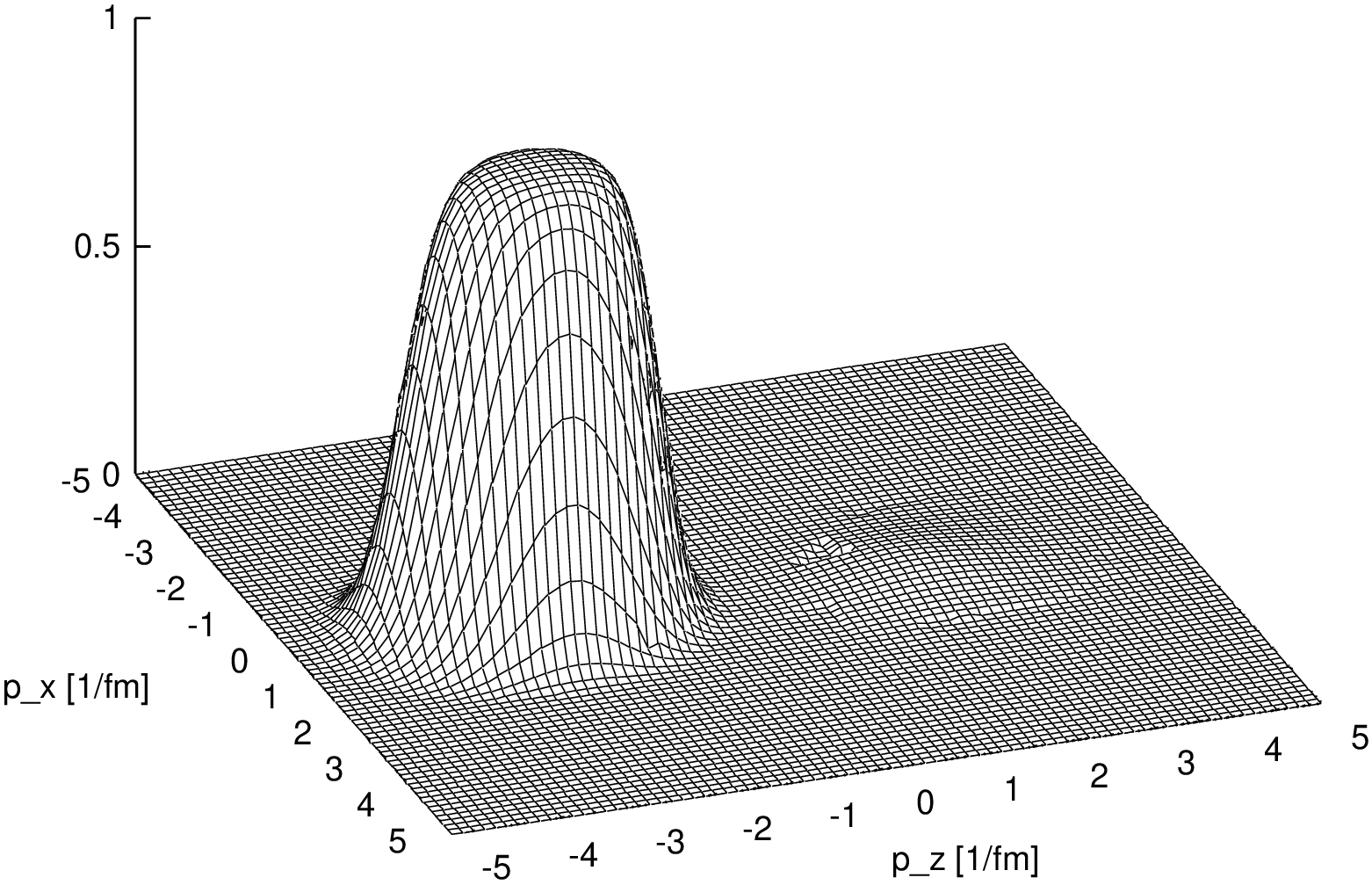,width=8.0cm}}}
\end{picture}
\caption{\label{fig1} Left: Local momentum distributions from RBUU at 
positions ${\bf x}=(x,y,z)=(0,0,0)$ (top), ${\bf x}=(0,0,2)$ (middle) 
and ${\bf x}=(0,0,4)$ (bottom) (in units of [fm]) at time $t=10$ fm/c. 
The figures on the right give the corresponding ACNM momentum distributions 
at finite temperatures $T_{1,2}$ with values of 
$T_{1,2}=29$ MeV (top), $T_{1}=40$ and $T_{2}=8$ MeV (middle) and $T_{1}=25$ und 
$T_{2}=1$ (bottom). The considered reaction is a central 
($b=0$ fm) Au+Au reaction at $E_{beam}=600$ AMeV beam energy. }
\end{center}
\end{figure}

How to go beyond the zero-temperature approximation is difficult to study.
The DB approach \cite{DB1,DB2} has been mainly applied for ground state nuclear matter 
at zero temperature. The few existing finite temperature calculations 
indicate a moderate $T$-dependence of the mean field at  
temperatures of a several MeV \cite{DB1}. Thus, one can expect that 
the approximation should be reliable for small 
thermal excitations with moderate momentum tails. 
In heavy ion collisions high thermal excitations with 
local temperatures of $T \sim 20-40$ MeV in the central region during the 
compression phase can be reached at intermediate energies, (see Fig.~\ref{fig1}),
 which makes this approximation less obvious. 
 However, we believe that temperature effects become important in a higher order 
correction and, in addition, complicate the calculation significantly.
The determination of a local temperature in heavy ion collisions, which is
necessary for such a treatment, requires a fit procedure 
at each space-time point. Thus, we do not consider a possible temperature 
dependence of the mean field here.

The assumption of a symmetric CNM approximation  
should be reliable for central collisions and high 
beam energies. However, as seen in Fig. 1 this is not the case especially if one 
goes away from the central cell. 
We expect the asymmetry of the CNM configuration to be particularly important at low
energies. At high energies, on the other side,  faster time scales and
a higher transparency of the system  - for peripheral reactions - leads rather to 
a separation between $2$- and $1$-Fermi sphere configurations. Thus, we extend 
the CNM model to asymmetric $2$-Fermi ellipsoid situations (called as 
Asymmetric Colliding Nuclear Matter (ACNM) in the following).

\section{Asymmetric colliding nuclear matter in the relativistic DB approach}

The treatment of non-equilibrium effects within the spirit of a CNM approximation 
has been extensively investigated in ref. \cite{sehn}. Here we extend this 
approach to Asymmetric Colliding Nuclear Matter (ACNM) by introducing an asymmetry 
in the Fermi momenta of the two counter-streaming systems. 
We describe the idea of the formalism here; more detail can be 
found in ref. \cite{sehn}. 

\begin{figure}[t]
\begin{center}
\unitlength1cm
\begin{picture}(9,6.5)
\put(-2.25,0){\makebox{\epsfig{file=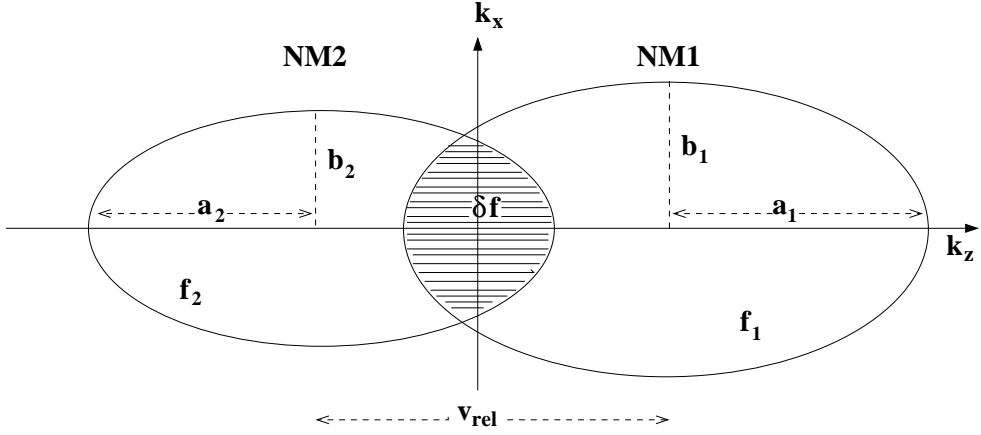,width=13.0cm}}}
\end{picture}
\caption{\label{fig2}Schematic representation of the ACNM configuration 
given by two covariant Fermi ellipsoidal momentum 
distributions $f_{{\rm i}}$, (i=1,2). These are characterized by Fermi momenta 
($b_{{\rm i}}=k_{F_{{\rm i}}}$) which are elongated along the boost direction 
($a_{{\rm i}}=\gamma_{{\rm i}} v_{{\rm i}} k_{F_{{\rm i}}}$).}
\end{center}
\end{figure}

The momentum distributions are described in terms of a superposition of two 
counter-streaming nuclear matter currents, i.e. two boosted Fermi ellipsoids at 
zero temperature. This configuration is schematically shown in Fig.~\ref{fig2}.
Analytically we write it as
\begin{eqnarray}
f_{12}({\bf k}) & = & f_{1}({\bf k},k_{F_{1}}) + f_{2}({\bf k},k_{F_{2}}) + 
                          \delta f({\bf k}) 
\nonumber\\
                    & = & \Theta(\mu^{*}_{1}-k_{\mu}u_{1}^{\mu}) + 
                          \Theta(\mu^{*}_{2}-k_{\mu}u_{2}^{\mu}) + 
                          \delta f({\bf k}^{*}) 
\label{f_acnm}
\qquad .
\end{eqnarray}
$\Theta$ is the step function, $k_{F_{{\rm i}}}$ (i=1,2) are the Fermi momenta and 
$u^{\nu}_{{\rm i}}=(\gamma_{{\rm i}},~{\bf u}_{{\rm i}}\gamma_{{\rm i}})$ 
are the streaming 
four velocities of the two boosted nuclear matter currents. The chemical 
potentials in the moving system are then given as
\begin{displaymath}
\mu_{{\rm i}}(k) = \mu_{{\rm i}}^{*}(k) + \Sigma_{\alpha}u^{\alpha}_{{\rm i}} = 
\sqrt{k_{F_{{\rm i}}}^{2}+m^{*2}} + \Sigma_{\alpha}u^{\alpha}_{{\rm i}}
\end{displaymath}
in terms of the effective chemical potentials $\mu_{{\rm i}}^{*}(k)$ 
in the rest system. 
The last term in eq. (\ref{f_acnm}) $\delta f({\bf k})$ guarantees 
the conservation of the Pauli principle in the overlap region between 
the two nuclear matter currents, and it is given by 
$\delta f({\bf k})=-\Theta_{1}({\bf k})\Theta_{2}({\bf k})$. 
However, to guarantee baryon number conservation one has to restore the total 
baryon density. This can be done in a covariant manner 
by redefining the Fermi momenta \cite{sehn}.

The ACNM configuration is characterized by three invariant parameters, 
the invariant densities of the two Fermi ellipsoids and the relative 
velocity 
\begin{equation}
\rho_{{\rm i}}=\sqrt{j_{{\rm i}\mu}j_{{\rm i}}^{\mu}} 
\quad\mbox{,}\mbox\quad 
v_{{\rm rel}}=\left| \frac{ {\bf v}_{1} - {\bf v}_{2} }{ 1 - {\bf v}_{1}{\bf v}_{2} }
\right|
\label{acnm_par}
\qquad .
\end{equation}
An alternate set of configuration parameters, denoted collectively by 
 $\chi$, are
$\chi = \{ v_{{\rm rel}}, \rho_{{\rm tot}}, \rho_\delta\}$
where 
$\rho_{{\rm tot}}=\rho_{1}+\rho_{2}$ is the total invariant density, and 
\begin{equation}
\rho_\delta=\frac{\rho_{1}-\rho_{2}}{\rho_{{\rm tot}}}
\end{equation}
the density asymmetry parameter. 
$\rho_\delta =0$ corresponds to the special cases of symmetric CNM, 
$\rho_\delta =1$ to a single Fermi-sphere at rest 
with invariant density $\rho_{1}=\rho_{{\rm tot}}$ and 
a single nucleon with relative velocity $v_{{\rm rel}}$. 
Thus, the ACNM model naturally includes a smooth transition to the 
local density approximation (LDA) of 
equilibrated nuclear matter for both cases $ v_{{\rm rel}}\mapsto 0$ and 
$\rho_\delta \mapsto 0$ while symmetric colliding nuclear matter contains 
only the first limit. 
All the expressions for the scalar and baryon densities can be derived in terms 
of these invariants. 

The relativistic self-energy in CNM as well as in ACNM has the same
Lorentz structure as in ground state matter, i.e. it contains a scalar part
entering into the effective mass $m^*$ and a vector part entering into
the kinetic four-momentum $k^{*}_\mu$ \cite{sehn,gait03} 
\begin{eqnarray}
\Sigma (k) &=& \Sigma_ {S}(k) - \gamma_\mu \Sigma^{\mu}(k)~~~.
\end{eqnarray}
The construction of ACNM is done in a manifestly covariant way and it can be defined 
in any reference frame. It is, however, most appropriate to work locally in
that frame where the total baryon current 
vanishes ${\bf j}_{12}={\bf j}_{1}+{\bf j}_{2}={\bf 0}$. We will denote it as 
current-zero frame $RS_{12}$ in the following and we will work always in $RS_{12}$. 
The space-like part of the vector field vanishes in $RS_{12}$ by definition 
${\bf \Sigma}_{12}={\bf \Sigma}_{1}+{\bf \Sigma}_{1}={\bf 0}$ and 
${\bf k}^{*}={\bf k}$. This simplifies the 
situation considerably, since canonical and kinetic momenta 
are not identical in general frames. 

As discussed above, we construct ACNM by extrapolating from calculations 
of ground state nuclear matter in the DB approach. The 
self-energies are thereby given as
\begin{equation}
\Sigma_{m} (k;k_F) = {\cal C} \: \int d^3q \: g_{m}(|{\bf q}|) \:
              f({\bf q}; k_F) \: {\cal T}_{m}(k,q) 
\label{sigma0}
\end{equation}
in terms of the distribution function $f({\bf q}; k_F)$ of one Fermi sphere 
(${\cal C} \equiv \frac{\kappa}{(2\pi)^{3}}$ with $\kappa=4$ for nuclear matter). 
Here $m=s,0$ stands for scalar and (time component) vector self-energies. 
The ${\cal T}$ matrix amplitudes can be taken from DB calculations and 
$  g_{m}$ in (\ref{sigma1}) are appropriate weight factors corresponding to the 
Lorentz structure of these amplitudes \cite{sehn,gait03}. 
The invariant ${\cal T}$ matrix amplitudes themselves are determined 
in equilibrated nuclear matter, i.e. for a single Fermi sphere. We further 
define effective coupling functions as
\begin{equation}
\Gamma_m = \Sigma_{m} \Big/ \rho_m \quad .
\label{gamma}
\end{equation} 
In Relativistic Mean Field (RMF) theory these quantities correspond 
to $g_m^2/m_m^2$, where $g_m$ are the coupling constants of the $m$-meson 
to the nucleon and $m_m$ its mass.

The ACNM self-energies are given in a corresponding way by using the ACNM 
distribution function of eq. (\ref{f_acnm}) for a configuration 
specified by the parameters $\chi = \{ v_{{\rm rel}}, \rho_{{\rm tot}}, \rho_\delta\}$ 
\begin{eqnarray}
\Sigma^{(12)}_{m} (k;\chi) & = & {\cal C} \: \int d^3q \: g_{m}(|{\bf q}|) \:
f_{12}({\bf q}; \chi) \: {\cal T}_{m}(k,q,\chi) \nonumber \\
 & = & \Sigma^{(1)}_{m}(k; \chi) + \Sigma^{(2)}_{m}(k; \chi) + \delta\Sigma_{m}(k; \chi)              
\label{sigma12}
\quad .
\end{eqnarray}
In the second line it is written 
 as a sum of contributions of two streaming Fermi spheres and a Pauli correction. 
Note, that the effective ${\cal T}$ matrix also depends on the configuration 
parameters, since the intermediate propagator does. In the first two terms 
the self energies for the 
two streaming nuclear matters are specified as (i=1,2)
\begin{eqnarray}
\Sigma^{({\rm i})}_{m}(k;\chi) & = & {\cal C} \: \int d^3q \: g_{m}(|{\bf q}|) \:
f_{{\rm i}}({\bf q}; \chi) \: {\cal T}_{m}(k,q,\chi) \nonumber \\
 & \simeq & \Gamma_m (\Lambda^{-1}k; k_F) \rho^{({\rm i})}_{m}(\chi)              
\label{sigma1}
\quad .
\end{eqnarray}
The second line contains the essential approximation of the approach, 
namely that the coupling function of ground state nuclear matter are used at 
momenta corresponding to the appropriate Lorentz boost multiplied
by the density of the ACNM configuration. The Pauli 
correction can be evaluated in a similar way. 
Since the fields are only moderately momentum 
dependent below the Fermi surface, the coupling functions 
can be determined by evaluating 
at the Fermi momentum. 
For the special case of symmetric matter 
(CNM, $k_{F_{1}}=k_{F_{2}} \equiv k_{F}$) one obtains 
\begin{eqnarray}
\delta\Sigma_{m}(k; \chi) & = & {\cal C} \: \int d^3q \: 
                   g_{m}(|{\bf q}|) \: 
                   {\cal T}_{m}(k,q) \: \delta f(q; \chi) 
                 = \Gamma_{m}(k_{F},k) \delta\rho_{m}(\chi) 
\nonumber\\
& \approx & \Gamma_{m}(k_{F},k_{F}) \delta\rho_{m}(\chi) 
\label{pauli-term}
\quad .
\end{eqnarray}

In this form all the self energies depend on the variables 
$\{k;\chi\}=\{k;v_{{\rm rel}},\rho_{{\rm tot}},\rho_{\delta}\}$. They 
are difficult to handle in a transport calculation, in particular 
because of the explicit momentum dependence. In analogy to the 
Hartree (or RMF) approximation we want to obtain momentum-independent 
self energies. These are obtained by averaging over the momentum
of the ACNM configuration
\begin{equation}
{\overline\Sigma}^{(12)}_{m} (\chi) = 
\frac{\int d^3k \: g_{m}(|{\bf q}|) \: \Sigma^{(12)}_{m}(k,\chi) \: f_{12}({\bf k},\chi)}   
     {\int d^3k \: g_{m}(|{\bf q}|) \: f_{12}({\bf k},\chi)} \quad 
     \equiv \quad {\overline\Gamma}^{(12)}_{m}(\chi) \: \rho_{m}^{(12)}(\chi)
     \label{sigmaav}
\quad ,
\end{equation}
where analogously as in eq. (\ref{gamma}) ACNM coupling functions 
are defined in the second equality.
Details of the averaging procedure and the calculation of the coupling 
functions are given in the Appendix. 

Fig.~\ref{fig3} shows the density asymmetry dependence of the effective coupling 
functions $\overline{\Gamma}_{s,0}^{(12)}$ (see Eq. (\ref{sigmaav})) 
at fixed relative velocity and total densities (solid circles). It is seen that 
the effective couplings rise with 
increasing asymmetry $\rho_{\delta}$ and approach  the LDA limit. 

At high relative velocities and 
moderate densities, where Pauli effects are only of minor importance, 
a rising asymmetry parameter shifts the center-of-mass towards 
the current with the higher density and the 
mean field is dominated by the momentum distribution of this ellipsoid. 
Thus, in averaging over the explicit momentum dependence of the 
fields, mainly moderate relative momenta contribute to the ACNM self energy. 
On the other hand, the invariants 
${\cal T}_{s,0}$ increase with decreasing momentum which leads finally 
to the observed $\rho_{\delta}$ dependence of the Fig.~\ref{fig3}. 
This behavior is more pronounced at larger densities since there 
the variation of the underlying T-matrix amplitudes is stronger.

The ACNM mean field depends on three parameters which 
complicates their application in transport calculations for 
heavy ion collisions. For practical use we 
apply therefore a simple parameterization of the form 
($U(\rho,v_{{\rm rel}},\delta)$ stands for scalar and vector self energies) 
\begin{equation}
U^{ACNM} \equiv U(\rho,v_{{\rm rel}},\rho_\delta) = 
\rho_{\delta}^{2} U^{LDA}(\rho)+(1-\rho_{\delta}^{2})U^{CNM}(\rho,v_{{\rm rel}})
\label{param}
\quad .
\end{equation}
This parameterization is also shown in Fig.~\ref{fig3} (solid lines). 
It is sufficiently accurate to be used in heavy ion collisions. 

\section{Application to heavy ion collisions}

For the theoretical description of heavy ion collisions we use 
the relativistic Boltzmann-Uehling-Uhlenbeck (RBUU) transport equation 
\cite{horror,gait01}. The mean field will be implemented in three 
approximations  as  discussed in the previous 
section: The symmetric CNM model has been studied in detail previously 
\cite{gait99,gait01}, its extension in the ACNM approach based on the 
parameterization (\ref{param}) is studied here for the first time. 
Results are also compared to the LDA where 
the mean field is only a function of the total density, i.e. 
taken at zero relative velocity $v_{{\rm rel}}=0$. 
The density and momentum dependence of the mean field is taken from the 
DB model of Ref. \cite{DB2}. The nuclear matter saturation 
properties obtained with the Bonn A potential 
are $\rho_{\rm sat} = 0.185$ fm$^{-3}$ and $E= -16.1$ MeV. 
With a compression modulus $K=230$ MeV one has a relatively soft EOS. 
In this context one should keep note that the consideration 
of the non-equilibrium effects in the CNM/ACNM approaches weakens 
the nucleon effective interaction leading to an even softer EOS 
than that given in the LDA. 
In all the simulations the collision integral is treated in the same way 
by using standard parameterizations for the elastic and inelastic total and 
differential cross sections \cite{cross} including $\Delta$ and $N^{*}$ resonances. 
The resonances are propagated in the same mean field as the nucleons 
and they decay into one and two-pion final channels. The pions 
are propagated under the influence of the Coulomb interaction. However, they interact 
strongly with the hadronic environment due to absorption. 

We have analyzed Au+Au collisions at intermediate 
energies in terms of collective flow effects given by its in-plane and out-of-plane 
components. They can be characterized by the first and second order Fourier 
coefficients of the azimuthal distributions,  
$N(\phi) \sim 1 + v_{1}cos(\phi) + 2v_{2}cos(2\phi) + \cdots$, 
and calculated as 
\begin{equation}
v_{1} = \left\langle\frac{p_{x}^{2}}{(p_{x}^{2}+p_{y}^{2})^{1/2}}
\right\rangle 
\quad\mbox{,}\quad
v_{2} = \left\langle\frac{p_{x}^{2}-p_{y}^{2}}{p_{x}^{2}+p_{y}^{2}}
\right\rangle 
\nonumber
\quad .
\end{equation}
In general, $v_{1,2}$ depend, apart from 
the beam energy and centrality,  on rapidity and transverse momentum. This has 
been studied extensively experimentally, 
e.g., in refs. \cite{fopir,pink}, and theoretically 
using non-relativistic approaches for the 
mean field in \cite{dani00,sahu98,hartnack98,bass98,v2bao,v2zheng}, 
and relativistically in 
\cite{gait99,gait01,koch} and \cite{v2dani} where in the latter case it was 
claimed to see a softening of the nuclear EOS as nuclear density becomes large. 

We begin the flow analysis by discussing the rapidity and transverse momentum 
dependence of the in-plane flow $v_{1}$ in the Figs.~\ref{fig4},\ref{fig5} 
for semi-central and peripheral Au+Au reactions at 0.25 and 0.4 AGeV. 
For a realistic simulation of the experimental conditions the RBUU events were 
passed through a phase space fragment coalescence algorithm 
including a reaction plane 
determination in the same way as the FOPI experiment \cite{crochet,fopinew}. 
The centrality selection was performed using the total multiplicity of charged 
particles, see e.g. \cite{gait01}. 

From these figures it can be seen that the non-equilibrium 
effects accounted by the ACNM and CNM models influence the directed 
in-plane flow only moderately. The differences are relatively biggest 
 at large transverse momenta $p_{t}^{(0)}$ and in the high rapidity regions. 
This is expected, since high $p_{t}^{(0)}$ particles are mainly emitted earlier 
when the matter is still anisotropic. Mostly, the in-plane flow 
with ACNM mean fields lies between the CNM and LDA results. 
This result is to be expected since ACNM interpolates the
self-energies between the two extreme cases,  CNM and LDA. 
The different flow pattern originate from the treatment of 
momentum anisotropies on the 
mean field level. The momentum dependence 
included in the CNM and ACNM approximation generally weakens 
the repulsive vector self energy, in particular at high incident 
energies, which leads to an effectively softer EOS compared to the nuclear matter 
EOS (LDA case) \cite{gait03}. This mechanism reduces the in-plane
flow. 

The differences become more pronounced for the elliptic flow shown in 
Figs.~\ref{fig6},\ref{fig7}. The elliptic flow is generally 
considered as a suitable tool to study the momentum dependence of 
the nuclear mean field \cite{pink,dani00,dani2}. 
The elliptic flow is created  during the high compression phase and 
governed by the pressure gradient due to the expansion of the initially 
compressed fireball as well as by shadowing effects 
of the spectator matter. Since the high density phase of the collision 
is still governed by local non-equilibrium \cite{gait03} one expects to see 
clear differences between the CNM, ACNM and LDA models. The model dependencies 
are most pronounced in the transverse momentum dependence of the 
elliptic flow (Fig.~\ref{fig6}). The energy dependence 
of the $p_{t}^{(0)}$ integrated elliptic flow 
(Fig.~\ref{fig7}) shows also clear signals from the highly anisotropic 
compression phase. The LDA yields a very strong elliptic flow and fails 
to describe the excitation function above 0.25 AGeV. The CNM 
approximation, on the other hand, reduces the 
flow, particularly, at high energies and provides a good fit to the data there, 
but is too small at low energies. 
At low energies the ACNM model lies between the two limiting cases, LDA and CNM,  
which results in a rather accurate description of the excitation 
function over the complete energy range from $0.1\div 4$ AGeV. 

The differences between LDA, CNM, and ACNM approaches 
have the same origin as for the in-plane flow although the 
model dependencies are now more pronounced. 
Accounting for  momentum anisotropies on the 
mean field level the repulsion of the model is weakened. 
Thus, due to the smaller pressure gradients the 
initially compressed system expands slower which effectively reduces 
the shadowing effect of spectator matter and the magnitude 
of the out-of-plane emission. The differences between CNM and 
ACNM can be understood from Fig.~\ref{fig3}. We also find that 
the asymmetry dependence is similar at other relative velocities and densities 
not shown here. The increase of the vector field with asymmetry parameter 
$\rho_{\delta}$ result in a stronger elliptic flow in ACNM as compared to 
CNM, but still smaller with respect to LDA, as expected. 
One obtains a smoother transition to LDA as the beam energy decreases. 

\section{Summary and conclusions}
We studied heavy ion collisions with a relativistic transport model and 
focused on non-equilibrium features of the phase space and their 
correlation to collective dynamics in terms of flow signals. 
The non-equilibrium effects were described by a Colliding Nuclear Matter model. 
We extended this model to two Fermi ellipsoids with 
an asymmetry in their densities, called as ACNM. 
The application of CNM and ACNM at intermediate energy heavy ion 
collisions was compared to a simple Local Density Approximation (LDA). 

We analyzed collective flow effects 
in and out of the reaction plane. It was found that the directed in-plane 
flow is only moderately affected by asymmetry effects in the ACNM approach, 
except if one considers the transverse momentum dependence at high $p_t$. 
For high energetic particles the ACNM model provided the best description 
of the data especially at the lowest bombarding energy considered 
($0.25$ AGeV). 

The elliptic flow excitation function 
is more sensitive to these different approaches since this observable is 
built during the early non-equilibrium stage of the collision. 
We observed a reduction of the elliptic flow in the ACNM and CNM models compared 
to LDA, where the ACNM model gave the best description of the data at 
intermediate energies. The asymmetry of the configuration was found to be 
particularly important at low energies and provided a smoother transition to the LDA. 

We interpreted the results by the fact that the repulsive vector field 
is essentially reduced in the CNM and ACNM cases which also decrease the 
magnitude of flows as compared to LDA. This considerably improves the comparison 
with flow data. The deviations between CNM and experiment especially at low energies 
in the elliptic flow excitation function seems to be resolved when 
asymmetric CNM configurations are considered (ACNM) where a smoother transition 
to the LDA case is now  observed. 

We conclude that non-equilibrium effects are important describing the 
collective dynamics of intermediate energy heavy ion collisions. A 
fully consistent treatment of heavy ion dynamics and nuclear structure 
equations within the framework of non-equilibrium transport 
theory is presently not possible. 
The CNM/ACNM models provide an approximative, but solvable tool 
to incorporate local non-equilibrium features more consistently at the 
level of the effective interaction. For definitive conclusions on the 
nuclear matter EOS it is important to study such approximations 
in complex systems. Our results can be considered as a first step in this direction. 
The next step would be to consider finite temperature effects, 
as well as the explicit momentum dependence of the fields.



\begin{figure}[t]
\unitlength1cm
\begin{picture}(8.,8.0)
\put(1.5,0.0){\makebox{\epsfig{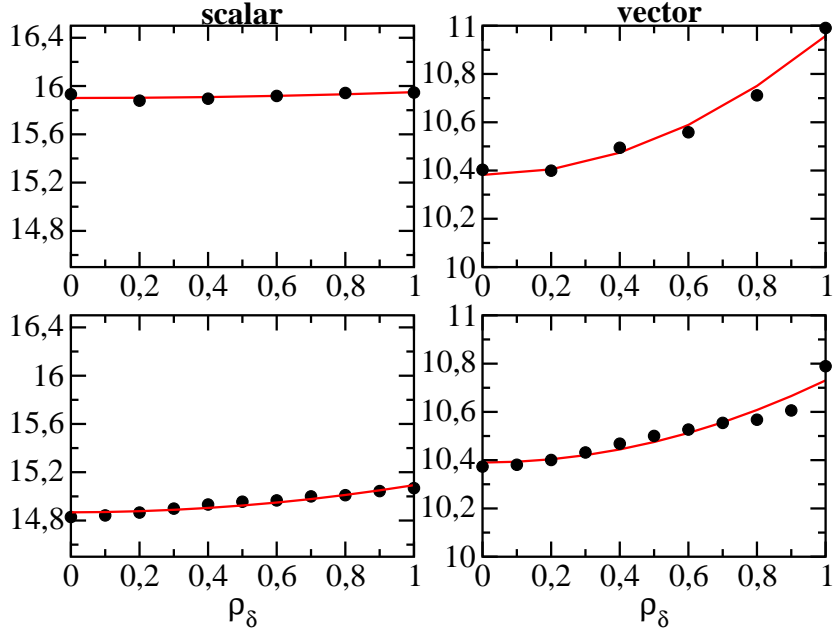}}}
\end{picture}
\caption{
Effective coupling functions $\overline{\Gamma}^{(12)}_{s,0}$ (left and right 
figures respectively) as function of the asymmetry parameter 
$\rho_{\delta}$ at fixed total densities of $\rho_{{\rm tot}}=1~[\rho_{{\rm sat}}]$ (top) and 
$\rho_{{\rm tot}}=2~[\rho_{{\rm sat}}]$ (bottom) and relative velocity $v_{{\rm rel}}=0.5$ [c]. 
The circles are results of ACNM calculations and 
the solid lines indicate the parameterization of Eq. (\protect\ref{param}).}
\label{fig3}
\end{figure}

\begin{figure}[t]
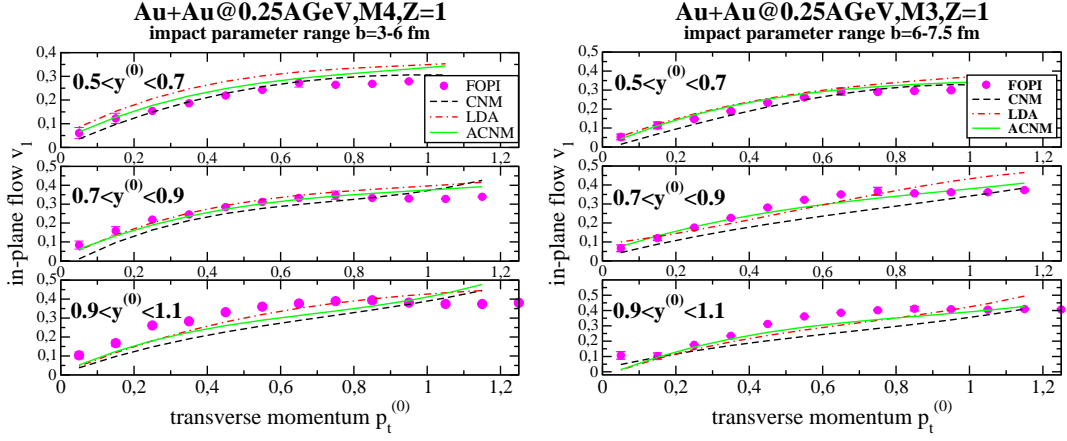

\unitlength1cm
\begin{picture}(8.,7.0)
\put(0.0,0.0){\makebox{\epsfig{file=fig4a.eps,width=6.9cm}}}
\put(7.2,0.0){\makebox{\epsfig{file=fig4b.eps,width=6.9cm}}}
\end{picture}
\caption{In-plane collective flow $v_{1}$ for charged ($Z=1$) particles 
in Au+Au reactions at 0.25 AGeV incident energy as function of the 
normalized transverse momentum 
$p_{t}^{(0)}$ at different rapidity intervals. 
The centrality intervals M4 and M3 corresponds to impact parameter 
ranges of $(3-6$ fm) and $(6-7.5$ fm), respectively. 
RBUU calculations treat the  mean field in the following 
approximations: local density approximation (LDA) 
where no configuration dependence is included (dot-dashed lines); 
colliding nuclear matter (CNM) approximation (dashed lines); 
asymmetric colliding nuclear matter (ACNM) approximation 
(solid lines). The FOPI data (filled circles) are taken 
from \protect\cite{fopinew}.}
\label{fig4}
\end{figure}

\begin{figure}[t]
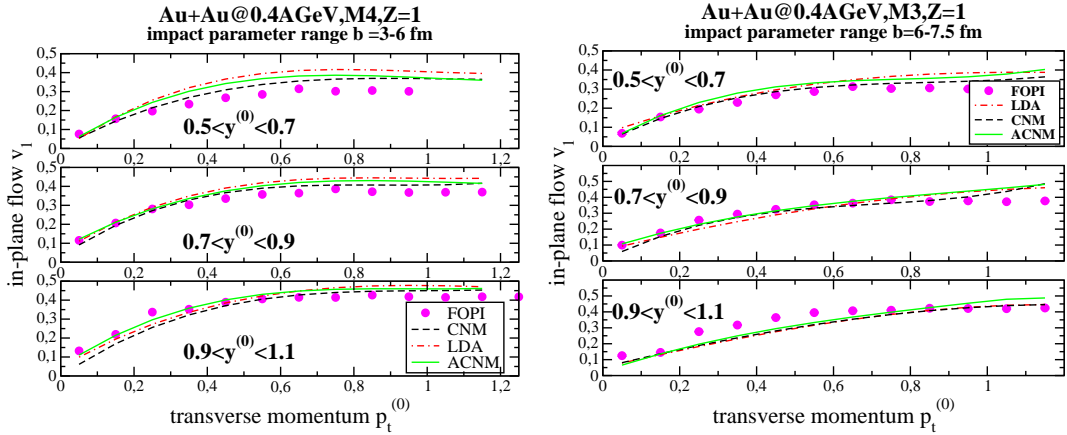

\unitlength1cm
\begin{picture}(8.,7.0)
\put(0.0,0.0){\makebox{\epsfig{file=fig5a.eps,width=6.9cm}}}
\put(7.2,0.0){\makebox{\epsfig{file=fig5b.eps,width=6.9cm}}}
\end{picture}
\caption{Same as in Fig.~\protect\ref{fig4}, but at 0.4 AGeV 
incident energy.
}
\label{fig5}
\end{figure}

\begin{figure}[t]
\unitlength1cm
\begin{picture}(8.,6.0)
\put(2.0,-0.5){\makebox{\epsfig{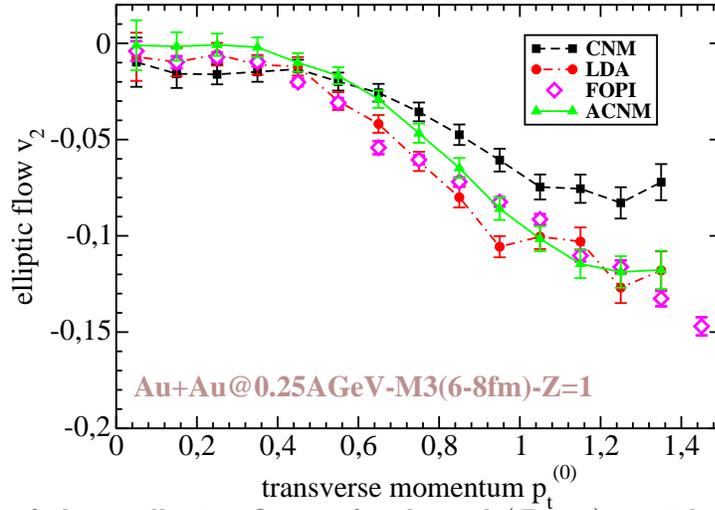}}}
\end{picture}
\caption{
Out-of-plane collective flow $v_{2}$ for charged ($Z=1$) particles 
as function of the normalized transverse momentum 
$p_{t}^{(0)}$ at mid rapidity $-0.15 \leq y^{(0)} \leq 0.15$ for 
the reactions as indicated. 
The models are the same as in Fig.~\protect\ref{fig4}. 
}
\label{fig6}
\end{figure}

\begin{figure}[t]
\unitlength1cm
\begin{picture}(8.,9.0)
\put(0.0,0.0){\makebox{\epsfig{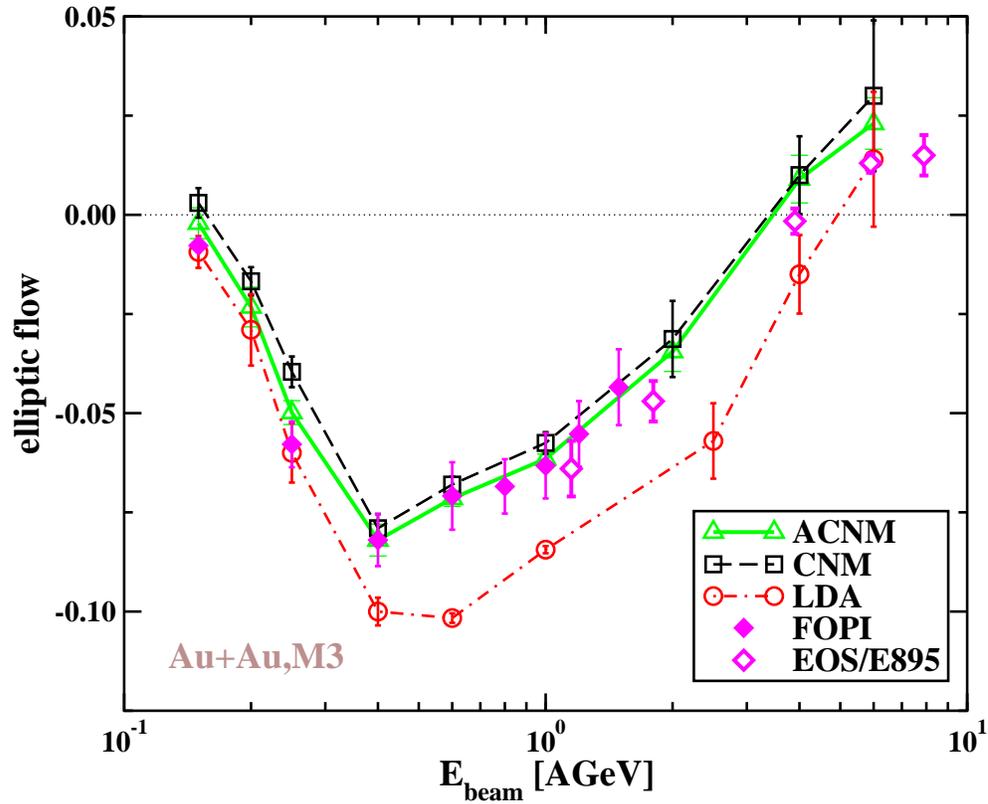}}}
\end{picture}
\caption{The excitation function of the elliptic flow $v_{2}$ 
in peripheral Au+Au collisions obtained with the models of 
Fig. \protect\ref{fig4}, see also \protect\cite{remark}. 
The data (open and filled diamonds) are taken from 
\protect\cite{pink,fopinew2}.
}
\label{fig7}
\end{figure}


\include{bib}

\include{anhang}

\end{document}

%% file: bib.tex

%% file: anhang.tex
\begin{appendix}
\section{Appendix}
\label{ACNM}
The ACNM configuration (\ref{f_acnm}) given by two counterstreaming 
nuclear matter currents is characterized by three parameteres, namely the 
Fermi momenta $k_{F_{1,2}}$, respectively the rest densities of the 
currents, and their relative velocity $v_{\rm rel}$. 
Expression (\ref{f_acnm}) contains a Pauli correction 
term $\delta f=-\Theta_{1}\Theta_{2}$ which restores the Pauli principle 
in case that the relative velocity of the currents is small and 
the original ellipsoids overlap. As discussed in detail in \cite{sehn} 
the Pauli correction can be performed in a covariant way. Corresponding 
expressions for the Pauli correction terms can be found in  \cite{sehn}.

In  \cite{sehn} the mean field, respectively the counterstreaming
self-energy components have been derived for symmetric  
nuclear matter configurations ($ k_{F_{1}}= k_{F_{2}}$). 
Here we present the extension to the general
case of asymmetric configurations ($ k_{F_{1}}\neq k_{F_{2}}$).
In order to obtain mean fields in a Hartree form which depend only on 
the ACNM parameters  $\{ k_{F_{1,2}}, v_{{\rm rel}}, \rho_{\delta}\}$ 
the self energy contributions of the two currents $\Sigma_{m}^{({\rm i})}$, 
(${\rm i}=1,2,\delta$, $m=s,0$) 
have to be averaged over the ACNM configuration in momentum space. This 
includes also corrections arising from the conservation of the Pauli
principle. 
 
\begin{figure}[t]
\unitlength1cm
\begin{picture}(14,5.5)
\put(2.5,0){\makebox{\psfig{file=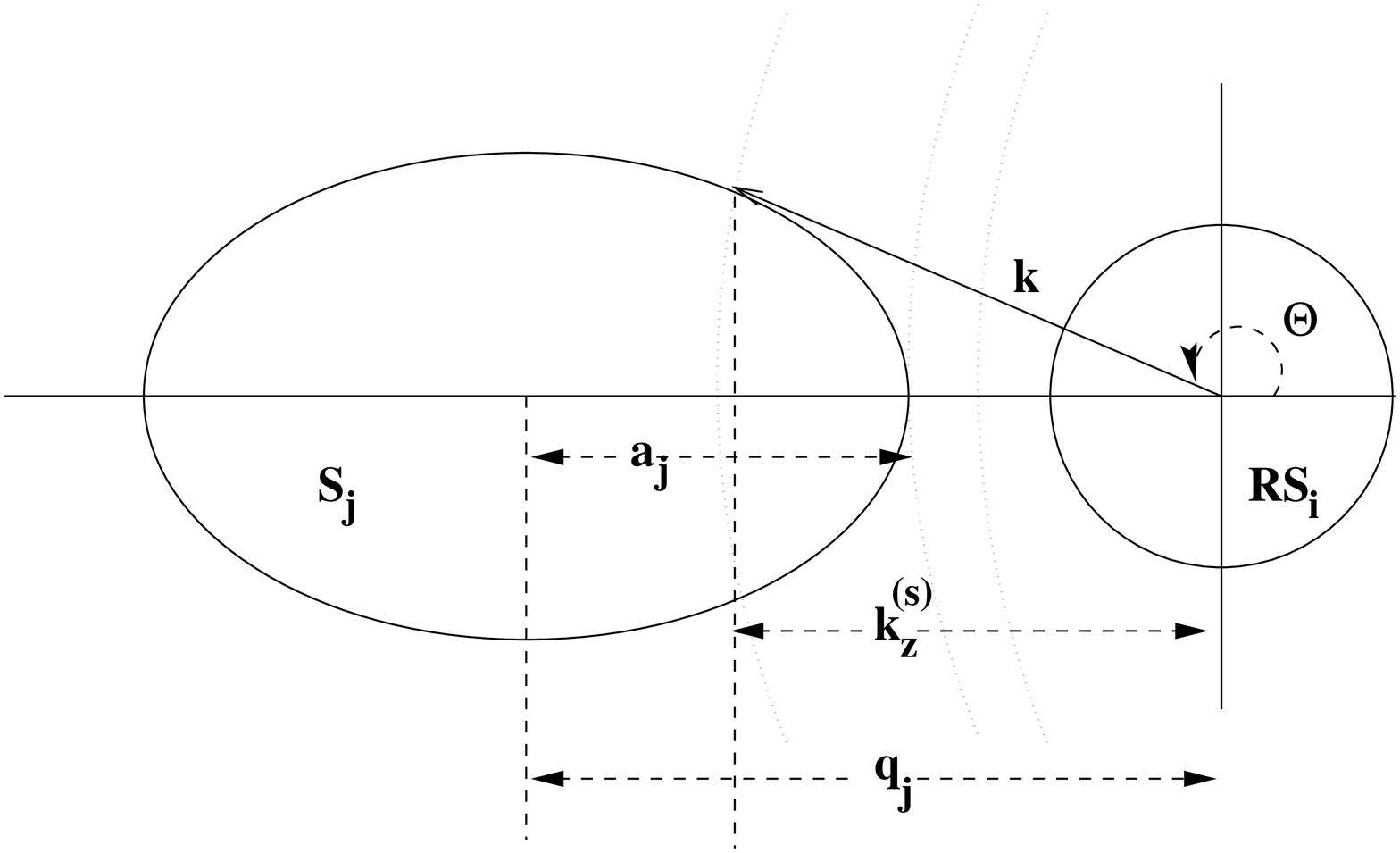,width=8.5cm}}}
\end{picture}
\caption{\label{fig8} Schematic representation of the averaging
of the  self energies 
over the ACNM distribution. 
The integration of $\Sigma_{m}^{({\rm i})}|_{RS_{{\rm i}}}$ has to be 
performed in the rest frame $RS_{{\rm i}}$ by 
integrating over the other ellipsoid in polar coordinates.}
\end{figure}

Since the underlying DB self-energies show only a moderate 
momentum dependence below the Fermi surface 
($k \leq k_{F}$) \cite{DB2} the integrations over the comoving current 
and also over the overlapp region are numerically easy to carry out due 
to moderate relative momenta. The difficulty appears for the remaining 
integrations of the self-energies over the 
second Fermi ellipsoid (${\rm i} \neq {\rm j}$)
\begin{displaymath}
\int \Sigma_{m}^{({\rm i})} f_{{\rm j}} 
\end{displaymath}
It is convinient 
to perform the integrations of the self energies 
$\Sigma_{m}^{({\rm i})}$ in their 
corresponding rest frames $RS_{{\rm i}}$. In this frame they depend only on the 
modulus of the $3$-momentum which reduces the integrations (in polar 
coordinates) to two dimensions. In particular, one has to perform 
integrations over spherical cells which include the integration area over 
the second ellipsoid as schematically depicted in Fig.~\ref{fig8}.


Let us for example consider the integration of $\Sigma_{s}^{(1)}|_{RS_{1}}$ over 
the other ellipsoid $S_{2}$, which has to be performed in the rest frame 
$RS_{1}$ (Fig.~\ref{fig8} with i=1, j=2):
\begin{displaymath}
{\cal C} \: \int d^3k \: \Gamma_{s1}(k_{F_{1}},k) f_{2} \: \frac{m^{*}}{E^{*}} 
= m^{*} {\cal C} \: \int_{k_{min}}^{k_{max}} k^{2}dk \: \int_{\Omega(k)} d\Omega(k) \: 
\Gamma_{s1}(k_{F_{1}},k) f_{2} \: \frac{m^{*}}{\sqrt{k^2+m^*}}
\qquad .
\end{displaymath}
The integration of the spatial angel $\cos\theta=\frac{k_{z}^{(s)}}{k}$ yields
\begin{displaymath}
\int_{\Omega(k)} d\Omega(k) = -2\pi \: \int_{k_{z}^{(s)}/k}^{-1} = 
2\pi \: ( -1-\frac{k_{z}^{(s)}}{k})
\qquad .
\end{displaymath}
The intersection point $k_{z}^{(s)}$ of the sphere with radius $k_{z}$ with the second 
ellipsoid $2$ is thereby given as \cite{sehn}
\begin{equation}
k_{z}^{(s)} = \frac{\gamma_{1}E_{F_{2}}-\gamma_{2} E_{F_{1}}}
                   {\gamma_{1}\gamma_{2}(v_{1}-v_{2})}
\label{schnittpunkt}
\qquad .
\end{equation}
Using eq. (\ref{schnittpunkt}) with $v_{1}=0$, $v_{2}=-v_{{\rm rel}}$, 
$q_{1}=0$, $\gamma_{1}=1$ und $k_{F_{1}}=k$ one obtains 
\begin{displaymath}
k_{z}^{(s)}|_{RS_{1}} = 
\frac{ E_{F_{2}}-\gamma(v_{{\rm rel}})\sqrt{k^{2}+m^{*2}} }
                             { \gamma(v_{{\rm rel}})v_{{\rm rel}} }
\qquad .
\end{displaymath}
In summary the integration over the polar angles leads to 
\begin{displaymath}
\Omega(k) = 
\left\{ \begin{array}{lr}
2\pi \big[ 1+\frac{1}{v_{{\rm rel}}k}\big(\frac{E_{F_{2}}}{\gamma}-
\sqrt{k^2+m^{*2}}\big) \big] & 
\mbox{in $RS_{1}$} \\
2\pi \big[ 1+\frac{1}{v_{{\rm rel}}k}\big(\frac{E_{F_{1}}}{\gamma}-
\sqrt{k^2+m^{*2}} \big) \big] &
\mbox{in $RS_{2}$} 
\end{array} \right\}
\qquad .
\end{displaymath}
In general, there exist different cases depending on the position of the center of 
the ellipsoid $S_{{\rm j}}$, i.e. it can lie inside or outside the sphere with 
radius $k$. This leads to different cases for the integration $\int d\Omega$:
\begin{eqnarray}
\mbox{$S_{{\rm j}}$ outside of $RS_{{\rm i}}$} & = & 
\left\{ \begin{array}{lr}
|q_{j}|-a_{j} < k < |q_{j}|+a_{2} & \int d\Omega(k)=2\pi(1+k_{z}^{(s)}/k) \\
          otherwise                   & \int d\Omega(k)=0     
\end{array} \right\}               
\nonumber\\
\mbox{$S_{j}$ inside $RS_{i}$} & = &
\left\{ \begin{array}{lr}
     0        \leq k \leq a_{j}-|q_{j}| & \int d\Omega(k)=4\pi \\
a_{j}-|q_{j}| <    k <    a_{j}+|q_{j}| & \int d\Omega(k)=2\pi(1+k_{z}^{(s)}/k) \\
                   k \geq a_{j}+|q_{j}| & \int d\Omega(k)=0
\end{array} \right\}
\qquad .
\label{diff}
\end{eqnarray}
The cases (\ref{diff}) can be summarized. For the scalar part of 
the self energy $\Sigma_{s}^{({\rm i})}$ the integration over the second ellipsoid 
$S_{{\rm j}}$ in the rest frame $RS_{{\rm i}}$ reads
\begin{eqnarray}
& & m^{*} {\cal C} \: \int_{k_{min}}^{k^{max}} k^{2}dk \: 
\frac{ \Gamma_{si}(k_{F_{i}},k) }{ \sqrt{k^{2}+m^{*2}} } \cdot 
\int_{\Omega(k)} d\Omega(k) = 
\nonumber\\
& & 2\pi m^{*} \: \Bigg\{ 
\int_{0}^{max\{0, a_{j}-|q_{j}|\}} 
\frac{ \Gamma_{si}(k_{F_{i}},k) }{ \sqrt{k^{2}+m^{*2}} } \: 2k^{2}dk + 
\nonumber\\
& & \int_{|a_{j}-|q_{j}||}^{a_{j}+|q_{j}|} 
\frac{ \Gamma_{si}(k_{F_{i}},k) }{ \sqrt{k^{2}+m^{*2}} } \: 
\bigg[ k^{2} + \frac{k}{v_{rel}}\left(\frac{E_{F_{j}}}{\gamma}-
\sqrt{k^{2}+m^{*2}} \right) \bigg] dk \Bigg\}
\nonumber
\qquad .
\end{eqnarray}
The total scalar part of the ACNM self energy averaged over the 
ACNM configuration reads finally
\begin{eqnarray}
& & \overline{\Sigma}^{(12)}_{s}(\chi) \: \rho_{s}^{(12)}(\chi) = 
\Bigg\{ \: \Gamma_{s1}(k_{F_{1}})\rho_{s}^{(1)2}(\chi) + 
\Gamma_{s2}(k_{F_{2}})\rho_{s}^{(2)2}(\chi) \: + 
\nonumber\\
& & 2\pi m^{*} {\cal C} \Bigg[ 
\int\limits_{0}^{max[0,\gamma(k_{F_{1}}-v_{rel}F_{F_{1}})]} 
\: \frac{ \Gamma_{s2}(k_{F_{2}},k) }{ \sqrt{k^2+m^{*2}} } \: 
2 k^2 dk \: +  
\nonumber\\
& & \hspace*{1.4cm} \int\limits_{\gamma | k_{F_{1}}-v_{rel}E_{F_{1}}| }^{\gamma \left(  k_{F_{1}}+v_{rel}E_{F_{1}} \right)} \: 
\frac{ \Gamma_{s2}(k_{F_{2}},k) }{ \sqrt{k^2+m^{*2}} } \: 
\Bigg( k^2 + \frac{k}{v_{rel}} \left( \frac{E_{F_{1}}}{\gamma}-\sqrt{k^2+m^{*2}} 
\right) \: \Bigg) dk \: \Bigg] \cdot \rho_{s}^{(2)}(\chi) \: + 
\nonumber\\
& & 2\pi m^{*} {\cal C} 
\Bigg[ \int\limits_{0}^{max[0,\gamma(k_{F_{2}}-v_{rel}F_{F_{2}})]} 
\: \frac{ \Gamma_{s1}(k_{F_{1}},k) }{ \sqrt{k^2+m^{*2}} } \: 
2 k^2 dk +
\nonumber\\  
& & \hspace*{1.4cm} \int\limits_{\gamma | k_{F_{2}}-v_{rel}E_{F_{2}} | }^{\gamma \left(  k_{F_{2}}+v_{rel}E_{F_{2}} \right)} \: 
\frac{ \Gamma_{s1}(k_{F_{1}},k) }{ \sqrt{k^2+m^{*2}} } \: 
\Bigg( k^2 + \frac{k}{v_{rel}} \left( \frac{E_{F_{2}}}{\gamma}-\sqrt{k^2+m^{*2}} 
\right) \: \Bigg) dk \: \Bigg] \cdot \rho_{s}^{(1)}(\chi) \: +
\nonumber\\
& & 2 \delta\rho_{s}(\chi) \sum_{j=1,2} \Gamma_{sj}(k_{F_{j}})\rho_{s}^{(j)}(\chi) + 
         \Gamma_{s}(min(k_{F_{1}},k_{F_{2}})) \delta\rho_{s}^{2}(\chi) 
\Bigg\}
\label{sigmas12_ACNM}
\end{eqnarray}
The scalar densities depend on the ACNM parameters $\chi$ due to the configuration 
dependence of the effective mass $m^{*}=m^{*}(\chi)$. 
The determination of the vector part of the mean field can be performed in a similar 
way. First one defines it covariantly by
\begin{eqnarray}
\overline{\Sigma}^{(12)\mu}(\chi) & = & {\cal C} \int d^3k \: \Sigma^{(12)\mu}(k;\chi) 
f_{12}(k;\chi) \frac{k^{\nu}}{E} j_{12\nu} / j_{12\alpha}j_{12}^{\alpha} 
\nonumber\\
 & = & < \Sigma^{(12)\mu}(k;\chi) f_{12}(k;\chi) \frac{k^{\nu}}{E} > 
j_{12\nu} / j_{12\alpha}j_{12}^{\alpha}
\nonumber\\
& \equiv & T^{\mu\nu}_{vec} j_{12\nu} / j_{12\alpha}j_{12}^{\alpha}
\label{mean_vector}
\qquad ,
\end{eqnarray}
with $T^{\mu\nu}_{vec} \equiv < \Sigma^{(12)\mu} f_{12} k^{\nu}/E >$. 
Eq. (\ref{mean_vector}) is valid in any reference frame, hence also in the special 
frame $RS12$ where the total baryon current vanishes:
\begin{eqnarray}
\overline{\Sigma}^{(12)\mu}
\equiv T^{\mu\nu}_{vec} j_{12\nu} / j_{12\alpha}j_{12}^{\alpha}
   & = & \Big[ 
         \Gamma_{01} j_{1}^{\mu} j_{1}^{\nu} + 
         \Gamma_{02} j_{2}^{\mu} j_{2}^{\nu} + 
         \overline{\Gamma}_{02}^{1} j_{1}^{\mu} j_{2}^{\nu} + 
         \overline{\Gamma}_{01}^{2} j_{2}^{\mu} j_{1}^{\nu} + \Big.
\nonumber\\
    & & \Big.  \Gamma_{01} \:\left( 
                     j_{1}^{\mu} \delta j^{\nu} + \delta j^{\mu} j_{1}^{\nu} 
                       \right) + 
         \Gamma_{02} \:\left( 
                     j_{2}^{\mu} \delta j^{\nu} + \delta j^{\mu} j_{2}^{\nu} 
                       \right) + \Big.
\nonumber\\
   & & \Big.  \Gamma_{0\delta} \delta j^{\mu} \delta j^{\nu}
        \Big] j_{12\nu} / j_{12\alpha}j_{12}^{\alpha}
\label{sigma012_ACNM}
\qquad .
\end{eqnarray}
Here $\Gamma_{0i}$ contains the integration over the comoving ellipsoid $RS_{i}$, 
and $\overline{\Gamma}_{0i}^{j}$ denotes the corresponding integration 
over the second ellipsoid $j$. These are exactly the same integrations as 
discussed above. 

From the averaged self energies (\ref{sigmas12_ACNM},\ref{sigma012_ACNM}) 
one can finally derive effective coupling functions for the ACNM configuration according 
to Eq. (\ref{gamma}) with $m \equiv s,0$
\begin{equation}
\overline{\Gamma}_{m}^{(12)}(\chi) = 
\frac{ \overline{\Sigma}^{(12)}_{m}(\chi)}{\rho_{m}^{(12)}(\chi)}
\label{gamma_acnm}
\quad .
\end{equation}
\end{appendix}

%% file: acnm_revised.bbl
\begin{thebibliography}{00}
\bibitem{remark}
We have repeated the calculations for the elliptic flow excitation 
function for peripheral Au+Au reactions up to 1 AGeV to achieve 
results with more precise statistics. The theoretical calculations 
of Fig. \protect\ref{fig7} with the LDA and CNM mean fields 
only slighly differ from those already published in 
\protect\cite{gait03}.
\end{thebibliography}
